\documentclass[oneside,twocolumn]{article}
\usepackage[utf8]{inputenc}
\usepackage{url}
\usepackage[backend=biber]{biblatex}
\usepackage{nopageno}
\usepackage{hyperref}
\usepackage{xcolor}
\usepackage{graphicx}
\usepackage{setspace}
\usepackage{enumerate}
\usepackage{array}
\usepackage{ragged2e}
\usepackage{amsmath}
\usepackage{appendix}
\usepackage{authblk}
\usepackage{mathtools}
\usepackage{mathptmx}
\usepackage{mathrsfs}
\usepackage{amssymb}
\usepackage{amsmath}
\usepackage{amssymb}
\usepackage{csquotes}

\begin{document}

\title{\huge{YASM (Yet Another Surveillance Mechanism)}}

\author[1]{Kaspar Rosager Ludvigsen}
\author[2]{Shishir Nagaraja} 
\author[3]{Angela Daly}

\affil[1]{Department of Computer and Information Sciences, University of Strathclyde, kaspar.rosager-ludvigsen@strath.ac.uk}
\affil[2]{Department of Computer and Information Sciences, University of Strathclyde, shishir.nagaraja@strath.ac.uk}
\affil[3]{Leverhulme Research Centre for Forensic Science and Dundee Law School, adaly001@dundee.ac.uk}

\maketitle

\begin{abstract}

Many types of surveillance exist on anything from smartphones to IoT devices, but most of them are not as ubiquitous and intrusive as Client Side Scanning (CSS) for Child Sexual Abuse Material Detection (CSAMD). Apple proposed to scan their software and hardware for such imagery. While CSAMD was since pushed back, the European Union has decided to propose forced CSS to combat and prevent child sexual abuse via a new regulation, and deliberately weaken encryption on all messaging services. CSS represents mass surveillance of personal property, in this case pictures and text, proposed by Apple without proper consideration of privacy, cybersecurity and legal consequences. We first argue why CSS should be limited or not used at all, and briefly discuss some clear issues with the way pictures cryptographically are handled and how the CSAMD claims to preserve privacy. Afterwards, in the second part, we analyse the possible human rights violations which CSS in general can cause within the regime of the European Convention on Human Rights. The focus is the harm which the system may cause to individuals, and we also comment on the proposed European Union Regulation. We find that CSS by itself is problematic because they can rarely fulfil the purposes which they are built for. This comes to down to how software is not  ``perfect'', as seen with antivirus software. Secondarily, the costs for attempting to solve issues such as CSAM far outweigh the benefits, and this is not likely to change regardless of how the technology develops. We furthermore find the CSAMD as proposed is not likely to preserve the privacy or security in the way of which it is described in Apple's own materials. We also find that the CSAMD system and CSS in general would likely violate the Right to a Fair Trial, Right to Privacy and Freedom of Expression. This is because the pictures could have been obtained in a way that could make any trial against a legitimate perpetrator inadmissible or violate their right for a fair trial, the lack of any safeguards to protect privacy on national legal level, which would violate the Right for Privacy, and it is unclear if the kind of scanning which would be done here could pass the legal test which Freedom of Expression requires, making it likely violate this as well. Finally, we find significant issues with the proposed Child Abuse Regulation. This is because it relies on techno-solutionist arguments without substance, disregards conventional knowledge on cybersecurity and does not justify the independence and power of a ``centre'' to help solve the problem.

%PLACEHOLDER ABSTRACT

%\keywords{Client-side Scanning \and Human Rights \and European Convention on Human Rights \and
%Encryption}
    
\end{abstract}

\section{Introduction}

%Easy to add something, hard to remove (the system), use historical CISCO firewalls as an example 
%\footnote{See for example \url{https://firstmonday.org/ojs/index.php/fm/article/download/1044/965?inline=1}.} Use these citations as examples from the past of the role firewalls played (and still do) as surveillance systems \cite{AndrewSchu}, \cite{Taillon2004}, \cite{Kim2006}, \cite{Johnson2007}

%Should include intro to CSAMD (Child Sexual Abuse Material Detection) but leave the technical details for below. Must have motivation which is how it affects a whole range of users of the same hardware and software. Furthermore, it is an example of the dismantling of existing privacy preserving techniques and replacing it with what is essentially systematic decryption. 

%Furthermore, it must include small introductions to each legal area. Maybe a bit about how this is a detection technique that can affect at least 3 legal spheres at once. 

%CSAMD was/is the testing of water for overt surveillance - which was not very socially acceptable seemingly.

%More sources will be added later 

%Yet another morality monitoring tool. 
%\cite[3]{Lauer2012}
Mass surveillance has been around for long \cite{Yekutieli2006}, and and a modern way to commit and complete the task now exists through digital mass surveillance \cite{Stoycheff2020}, but recklessly implementing and using this without proper cybersecurity (security) considerations can have devastating consequences, like the destruction of privacy through backdoors which criminals can use as easily as the state, or significant economic consequences \cite{Abelson2015}. Furthermore, using these digital tools do not seem to decrease complexity for the authorities \cite{Fussey2022}. Opponents argue that surveillance is necessary and benefits from secrecy \cite{Kolaszynski2022}, while basic cybersecurity engineering practices speak directly against this \cite{Saltzer1975, Anderson2020}. With the advent of a fully digital society \cite{Jørgensen2021}, surveillance is conducted on almost all hardware and software, and this increased during the COVID-19 crisis via digital health surveillance \cite{Sekalala2020, datajustice2020, Lyon2021}. 

Our focus of this paper is \emph{Client-side scanning (CSS)}, which is an example of a digital tool used for surveillance. The cybersecurity community broadly distinguishes between two types of scanning; server-side and CSS. CSS used on individuals has in particular been heavily criticized \cite{Abelson2021}. Most current surveillance is done in the former format, while existing software such as antivirus programs do the latter. Furthermore, CSS analyses the content before it is encrypted, unlike server-side, which creates many additional cybersecurity risks and potential failures. CSS therefore adds a whole new level of surveillance; these systems track everything in a given area real-time. Analogy wise, it will be like going from an agent occasionally tracking you and your house, to the agent living in your house and constantly going through your belongings. Types of malware have and still use what is essentially CSS to achieve their objectives \cite{Or-Meir2019}. Alongside this, the enabling of digital surveillance is continuously expanded on a worldwide scale. T

his can be seen in a country like the UK, through the stifling of protests, inability to sue the state \cite{Siddique2021} and disable certain groups from participating in a democratic society. Whether or not these measures become reality matters little, when the purpose of them clearly is to increase the opportunities to do surveillance and decrease the amount of rights for individuals. Such acts has been met with increased usage of encryption and other evasive measures, by everything from ``unwanted parties'' \cite{Steel2020} to dissidents, making national issues into international. But having encryption seems to be worth far more than not having it \cite{Abelson2015, Mann2020}. 

In this paper, we analyse whether CSS will violate human rights through the European Convention on Human Rights (ECHR) \cite{ECHR}. In it, The Right to Remain Silent and Not Incriminate Oneself can lead to failed prosecutions of potential criminals and a widespread waste of resources if violated. Surveillance systems are well known to violate rights, but CSS present systems which will do this routinely or constantly, which is why we find them to be dangerous and cannot justify by the goals they aim to serve. The same applies to antivirus, which can never fully eradicate all malware \cite{Or-Meir2019}. 

CSS will do this and also likely violate a myriad of human rights. CSS will likely be used regardless, which is why this paper includes comments on both technical and human rights issues. Usage of digital mass surveillance requires access to systems, done through lawful authorisation or outright illegal means \cite{Leman-Langlois2018}. We have seen a move to large corporations performing surveillance in the platform economy, which either enables them to act as gatekeepers or collaborators with national states \cite{Khan2018}, or lets them act entirely for their own benefit \cite{West2019}. Both include snooping on sounds, intercepting messages on the software they own and run, infer or read location data and access to personal files or pictures. These mirror the past types of physical surveillance. 

CSS can be exemplified by Apple's Child Sexual Abuse Material detection system (CSAMD). This system is postponed, but this does not make the discussion of it and future similar systems any less important. It would have applied to all IPhones and ICloud. Apple is not the only company to have such a system, but they are, as of the time of writing, the only company to have publicly released proofs and system summaries \cite{Bhowmick2021}. CSAMD was created to fight CSAM, which is broadly considered a crime in a majority of jurisdictions \cite{Nair2007}. Fighting such material is vital, but this is usually left to authorities of individual national states or cooperation in legal entities such as the European Union, Europol or Interpol. Outside of criminal law,  mental health surveillance is fought through apps \cite{Cosgrove2020} or through traditional means of data gathering \cite{Johnson2020}, but CSS goes beyond these. While the spread of CSAM is a major issue, preventing or mitigating the actions leading to the material being created is much more vital \cite{Assini-Meytin2020}, and is not something any private entity can ever do by itself. An additional danger which CSS brings in the field of CSAM detection and management, is to prematurely exhaust police resources, analogous to the damage which sensor-based monitoring systems did to chemical plants in the past \cite{Leveson1995, Kletz2009}. 

Regrettably, there is no doubt that states in the future will use CSS systems that apply the same kind of techniques as the CSAMD to other situations. This will likely not contribute to more security or more privacy\cite{Abelson2021}, and is not a new or unexpected development \cite{Rosenzweig2020, Huang2004, Shahriar2013} and have been used by adversaries in various forms for decades through malware \cite{Stone-Gross2013, Prakash2016}. The European Union has joined this list, with its new proposal for legislation, which frames CSS as the solution to all problems regarding child sexual abuse, and may mandate backdoors in all encryption on all communication platforms going forward, see Section 5 on page \pageref{Section 5}. CSS and data protection and privacy is a known angle \cite{Cobbe2021}, but research on the rights of individuals is needed. Since CSS will affect almost everyone, we chose to focus on human rights as they are laid out via the ECHR, as it covers the EU and more. 

The paper is as follows: Section 2 includes arguments against CSS, section 3 defines terms used in the paper, section 4 analyses human rights, with the case being ECHR, specifically Article 6, 8, 10 and 11. Section 5 discusses the proposed Regulation, section 6 includes future considerations and section 7 is the conclusion. 

We find that CSS has serious issues on a fundamental level, which cannot be solved since they exist because of their designs. Furthermore, we find that Apple's CSAMD was inadequate both in terms of its privacy preserving functions, NeuralHashing and even in its premise. It will lead to future CSS systems which can use other datasets and for other purposes, such as catching dissidents or protesters. To that extent, we find that CSS systems will violate several rights within the European Convention of Human Rights, but our analysis is not exhaustive. They will likely violate the Right to a Fair Trial, in particular the Right to Remain Silent and Not Incriminate Oneself, Right to Privacy, and if implemented further than current examples, Freedom of Assembly and Association as well. 

%They will likely pose further issues specifically in criminal law in most jurisdictions, as they can lead to wrongful prosecution and investigation or lead to situations where the user of manufacturer of the system can be kept liable.

%Include Human Rights, Democracy, and Rule of Law Impact Assessment ideas from https://datasociety.net/announcements/2021/11/22/recommendations-for-incorporating-human-rights-into-ai-impact-assessments

\section{CSS as an Impossibility}\label{CSS impossible}

It is a widely held belief that CSS with the right safeguards and privacy enhancing technologies can be done ``properly'', while others have concluded that they can never be secure \cite[34]{Abelson2021} or justifiable in their current form \cite{Rosenzweig2020}, and calling CSS to be impossible is not far off. Not impossible in its practical implementation, but the impossibility to genuinely and rationally live up to its own spirit. We first present a simple proof to make that point, then we continue will an alternative way to show the same.

\subsection{Proof 1}\label{Proof 1} We make two assumptions: \emph{1.} Assume that there exists a perfect CSS technique and that this is implemented into a system, and it could, e.g., be used to detect CSAM.

\emph{2.} For CSAM, there is no perfect definition of CSAM which can be expressed as a general rule to identify it as a type of image. \label{Assumption 2} There exists legal definitions, but these do not constitute the rules which a CSS system can make decisions from, such as Article 34 in the United Nations Convention on the Rights of the Child 1989 and the Convention on the Protection of Children against Sexual Exploitation and Sexual Abuse. An adversary will upload an image which is scanned by the CSS. If the CSS detects that this is CSAM, the adversary will delete the image, and discontinue their usage of the service. By doing so, the CSS will never detect any CSAM from this adversary ever again, and therefore become redundant. This means, that no matter how perfect you make the detection of CSAM or other changes, the CSS is not capable of fully fulfilling its purposes, as any adversary can entirely circumvent its entire existence by using other avenues. 

Assumption 1 has the same flaw as the idea of a 'perfect antivirus program' \cite{Or-Meir2019} \cite{Cohen1987}. 

Beyond CSAM, the same circumvention is possible. Because of this, CSS will in most cases not be able to fulfill the purposes for which it was created for, and at best fails, but will at worst be created and continue to function for entirely different purposes like surveillance-capitalism.

\subsection{Proof 2} We make the following assumptions about such systems:
\emph{1.} The aims of a system must be able to reach their goals, idealised or not.

\emph{2.} The proposed goals must not conflict directly with the capabilities of a system.

\emph{3.} There is no such thing as a perfect defence.

\emph{4.} The risk of circumvention cannot be fully mitigated.

These three assumptions cover purposes for the system, lack of justifications for its goals and therefore its existence, and its impossibility. CSS contains in its very notion constant surveillance upon the system, and unlike pure logging, attempts to oversee all events within a given framework. This makes it very similar to software like antivirus, which we cannot be ``perfect'' \cite{chess2000, Or-Meir2019, Cohen1987} as the definition of malicious software can never define all the types in existence. CSS is also malicious against the user \cite{Abelson2021}. Therefore, CSS is impossible because its existence is adversarial to the user, regardless of whether they violate what the CSS searches for or not, as CSS by its existence creates a security and a human rights' hazard. Two wrongs do not make a right. Essentially, the idea of CSS as a positive or constructive force does not exist because of its function. The goal of these systems is usually to aid in investigations or otherwise provide information, but by doing so, CSS systems inherently violate their own goals. They commit to surveillance, which can be misused by everyone that can guess keys or otherwise attack the system, creating the same problems as they are trying to solve. The goals of any CSS will therefore be in conflict with its capabilities at all times.  Furthermore, there is no way to guarantee against misuse and therefore violation of the very purpose of the systems. One must assume that there will be a certain risk of adversarial failures, and that there is no way to ever mitigate everything. 
If there is always a risk, constantly scanning what is done live on a system by itself means that CSS acts more like a weakness, than in does as a tool to fulfill its goals. All three arguments combined show that the held beliefs and arguments for CSS do not hold up when faced with the aims, goals and capabilities of the system. This will not change going forward, regardless of technical advancements.

\section{Client-side Scanning} CSS is on-device analysis of data \cite{Abelson2021}, which is different from analysis from information gathered elsewhere. We will go through one example of CSS.

%malware and antimalware are examples of CSS, but we focus on scanning on encrypted data. We might find papers on embedded malware - but do any do it over encrypted files?

%\subsection{Location Based}

\subsection{CSAMD}

Apple documents its proposed system in a variety of online documents \cite{AppleInc2021, Bhowmick2021, Bellare2021}. There are direct issues with how it preserves privacy. The system does this through hashing and PSI, and theoretically only lets humans-in-the-loop access them after certain measures. But not only is this construction vulnerable to adversarial attackers, the way they ``identify'' CSAM could lead to scores of false positives or other issues. Since the subject of CSAM is grave, it is prudent to make a safe and secure system that does not potentially label thousands of individuals for possessing materials they do not in reality. 

A less overarching but more detailed analysis of the system has been done elsewhere \cite{Pinkas2021a}, its NeuralHash image tool has been reverse engineered several times \cite{Struppek2021}, and another image tool akin to it has also been reverse engineered and found vulnerable \cite{Athalye2021}. 

The proposed detection system is based on a range of techniques: private set intersection (PSI) \cite{Rinda2017}, Cockoo tables \cite{Pagh2001}, Secret sharing, Diffie-Helman problems \cite{Pan2021} and hardness, Naor-Reingold Diffie-Hellman \cite{Mefenza2020} random self reducibility, Coppersmith and Sudan Algorithm, Interleaved Reed-Solomon code under random noise and tuples. The idea of private set intersection refers a method primary set theory properties like $ A \cup B $. 

The CSAMD uses secret sharing to find this intersection, which is done between the user, Apple and the system. It is evident that letting Apple control the system tool which one relies on for secret sharing is problematic. You do not want to share the real intersection, so you will use $ A' \cup B'$ instead. 

The CSAMD uses Shamir Secret Sharing, which means that the secret is an element of a finite field ($s \in \mathbb{F} $) , of which a polynomial with a certain degree is picked. Because of the qualities of the finite field, shares of the secret can be created which rely on the polynomial, and then there exists a threshold for which the polynomial can be guessed. 

But if the system cannot get the necessary amount of shares to reach the threshold, only the user and Apple will know the secret. In the CSAMD, this is supposedly further modified so that not even the system nor Apple will know through synthetic shares, which are periodically uploaded by the user \cite{AppleInc2021}. But will the system know when it is handling synthetic or real matches? As the data sent by the users is put into cuckoo tables to hash the data. Cuckoo hashing uses two functions with a set size and two tables, where the key (in this case) can exist in both tables but never both at the same time \cite{Pagh2001}. Each function will be sent every time, regardless of whether the key is in one or the other. In this context, Diffie-Hellman hardness is used to assert how negligible the success of an adversarial attacker in guessing different elements of the tuple and a property of the tuples is, and Diffie-Hellman random self-reducibility is a means to use the fact that Diffie-Hellman tuples are not random tuples \cite{Pinkas2021a}. The latter is a core part of the creation of the key mentioned above. The shares (central to PSI) which are correct among the synthetic ones can be detected with the interleaved Reed-Solomon under random noise algorithm \cite{Bleichenbacher2003, Pinkas2021a}, which relies on the relationship between the values, and these are kept under control by always including a version of the picture in lower resolution, limiting the size to one which the algorithm satisfies its parameters. 

% @Shishir, finish this section and the rest of technical sections please.

%\subsubsection{Next layers in the system}

%add more

%\subsubsection{Transparency of the training dataset and other features}

%Anything can be put into the training sets. How does Apple prove their only PSI is happening and the base images are only CSAM? Is there transparency in the training set.

%\subsection{Future Systems}

%Ideas for future systems, preferably and categories for types of on-device control. This could be political control, personal control each with different methods. This could be framing for crimes for personal but also logging which people they have taken pictures of or with. Location control could be its own category? 

%Future CSS systems will likely expand their areas of control. We suggest that they will be:

%\begin{enumerate}
    %\item Political Control
    %\item Personal Control
    %\item Location Control
%\end{enumerate}

%Shishir can add something here

%\subsubsection{Risks of usage of third parties} 

%Adversarial comments, subsubsection could be changed namewise

%to be deleted

\subsubsection{Safety Comments and Criticism} There has been public criticism by prominent authors of the system \cite{Mayer2021a}, explicitly of NeuralHash \cite{Struppek2021} and similar future schemes \cite{Abelson2021}. 

Some developers \cite{Mayer2021a} still supports this type of surveillance, but such systems have similar human rights issues as \hyperref[4.0]{below}. 

CSAMD may fail to detect CSAM, because it is not within its known data set, adversary abuses NeuralHashing or by other measures. This could be embedded within a picture. This can be done both to abuse the system (to i.e. target individuals) or to hide CSAM as such \cite{Struppek2021}. \hyperref[Assumption 2]{Assumption 2} from Proof 1 still applies here, there is no perfect way to identify and otherwise always find CSAM. 

Reversely, any machine or other learning can therefore be abused or circumvented. CSAMD is a system that is likely made for the benefit of Apple, regardless of its otherwise altruistic intentions that clash with existing views \cite{Sanchez2019}. There a very real risk for false positives with potential consequences for the physical person targeted, but its choice of PETs is questionable and the risk of many types of adversarial failures is great. 

Struppek et al. \cite{Struppek2021} present compelling arguments which show that NeuralHash is not fit for purpose. 90 to 100 percent success rate \cite{Struppek2021} of an adversarial attack, show that NeuralHash is not robust against simple image processing software. This may indicate problems with deep perceptual hashing employed in this manner in general, and not just one images. Struppek et al. also showed that NeuralHash leaks from its classifiers, which is not unexpected, but further justifies not using any technique similar to NeuralHash when handling potentially sensitive or private information, as many adversaries will be able to infer information or perhaps more without having access to the entire image. The use of Neural Networks will always enable the risk of the attacks mentioned by Struppek et al., which means they should not be used until suitable defences are found. We agree, but philosophical arguments are not built around conclusive statements,  ``not fit for purpose'' does not argue strongly enough against using such systems when they seem this unfit. Future authors will give more cohesive and rigorous arguments as to why CSS can be unethical within defined ethical frameworks. 

A key to this is cause and effect, which through the lens of applied ethics like consequentialism \cite{Card2020} makes CSS hard to justify. The increased surveillance must equally increase safety, and must cause more efficient sanctioning and perhaps increased justice, but none of this is likely to happen \cite{Abelson2021}.

Finally, there is no point in discussing ``perfect'' privacy, if the supposed materials are later de-encrypted and then decided on by a human in the loop who is not part of law enforcement of any state. This is regardless of the supposed safeguards which NGOs \cite{Wagner2014} and other structures that may include specialists, as they will still not be faced with public ethical and legal rules of conduct or enjoy the legitimacy of being hired by the state in question. 

%this is something that can be written easily later

%section{Data Protection and Privacy}

%Such a decryption system violates GDPR within the means of privacy and handling of personal data (outside of privacy). Other issues can be handled by other authors, such as legal processing grounds etc. 1, there is a real risk of leakage, risking the privacy of every user and of every kind of image they may have on their devices. 2, the security of the system must fit the vague standards in the GDPR, and it seems to be under or worse than what they had before. This does not fit a Company the size of Apple, especially since this openly leads to less privacy regardless of the claims they have in their technical summary.

%subsection{Data Protection}

%GDPR - Preamble 39, 49, 75 (includes reputation damage), 78 (privacy and data protection by default), 83 (encryption etc.), 91 assessments (high risk and massive data processing), 94 (high risk from DPIA), Art. 2(12) (personal data breach), art. 5(1)(f), Art. 32 (general security article), 
%Police Directive, Directive 2016/680 applies. Not sure how relevant it is, maybe we should take a look. 

%subsection{Privacy}

%the Eprivacy Directive is the main source here. We may include commentary on how it may play into the Eprivacy Regulation.

\section{Human and Constitutional Rights} \label{4.0} CSS systems bring up human rights issues. Privacy violations seem clear from its inception and contextual privacy and clearer consent rules will not solve these issues \cite{Gharib2022}. The potential which systems inspired by the CSAMD have in the future present a violations of other rights. 

To some, questions of human rights or rights derived from constitutions seem \emph{unrealistic}, \emph{abstract}, or perhaps even \emph{redundant}. \emph{Unrealistic}, because the individual or even a given company may never reach the procedural point where it must be proven or refuted in court \cite{Klug2006}. \emph{Abstract}, because the concepts seemingly have nothing to do with the private or professional lives of those that have this belief, but this is misleading \cite{Ziebertz2018}. And \emph{redundant}, because it may seem like those rights are built in or part of the human nature, which does not seem to be so \cite{Donnelly1982}. The last century gave us the basis of these rights because of the atrocities that were committed to prevent further deprivation of these rights, even with cultural differences \cite{Chemhuru2018}.

The ECHR is an instrument was created by the Council of Europe, which includes EU Member States, and additional countries. The EU has the Charter of Fundamental Rights of the European Union, which uses and expands the ECHR. The ECHR exists above national legal systems, which is necessary for it to function \cite{Besson2011}. For a citizen in these countries, the European Court of Human Rights (ECtHR) is the last resort. Before going to the ECtHR, it is up to the national judges, civil or otherwise, to apply it properly. Not every case can be appealed to the ECtHR, as its case law has already answered the common questions. It is within this substance that judges and the systems as such must make these rights into reality. Human rights exist at all times, but can only be negatively or positively accessed in court. ``Negatively refers'' to what should not be done or inaction, and ``positively'' refers to what has to explicitly be done or acted upon. Any comments we make or analysis of how the CSAMD system may work, are in the context of hypothetical issues in the future. This does not mean it will only apply to the CSAMD, it will apply to any kind of automated mass personal data CSS analysis tool that any state may use in the future via function creep \cite{Koops2021}. 

The following section is therefore a direct analysis as to \emph{when} CSS systems can breach the ECHR and how. Our analysis cannot amount to a justification as to whether they should ever be used for these purposes, but serves an evaluation of legality of the systems within each article. 

\subsection{Right to a Fair Trial} \label{4.1} CSS may impact the Right to a Fair Trial (Article 6), if evidence gathered by CSS is used in court or during investigation. Many content moderating companies have internal or external measures to share with law authorities, but this may not be clear in all jurisdictions, and regarding CSAMD, Apple has not made it clear how it would complete its duties on a world wide scale. Article 6 applies if the individual is ``charged with a criminal offense''. Charged does not only refer to when criminal proceedings have begun, but also before or when the suspicion is internal.\footnote{31816/08, Stirmanov v. Russia, § 39.} 

%If CSS is used in cooperation with authorities at any point in future, it may trigger Article 6 immediately. 

\subsubsection{The Right to Remain Silent and Not Incriminate Oneself}\label{4.1.1} The Right to Remain Silent and Not Incriminate Oneself is not mentioned explicitly in Article 6, but is accepted through case law\footnote{10828/84, Funke v. France, § 44.}. 

The right implies that suspects have the right not to speak and not give information which would incriminate them, and applies outside of criminal law too \cite{Hupli2018}. The negative limit is to prevent evidence obtained through coercion or oppression\footnote{19187/91, Saunders v. the United Kingdom, § 68 - 69} from being used. The exception is where the evidence is obtained through compulsory powers (e.g., lawful authorisation), of which the evidence must have an independent existence of the subject (e.g. blood). 

CSAMD makes use of pictures which do not have an independent existence of the subject. Whether this applies to the NeuralHash which CSAMD has set up to not have a human in the loop until later, remains to be seen, and cannot be answered from existing verdicts. 

Many jurisdictions likely violate self incrimination through current practice like the analysis of social media posts of potential refugees in the application process \cite{Leurs2017}. Pictures which are synonymous with private property would also fit this category, denying its independent existence of the subject. 

If lawful authorisations are not systematically given access CSS systems, a test must be done by national courts to consider whether the individual retains the right: 

\emph{The nature and the degree of the compulsion, the existence of relevant safeguards in the procedure, and the use of the obtained material.} 

It is unlikely that the retrieved files or data can be the sole evidence using to convict the individual, regardless of the kind of comprehensive scanning client-side based systems can do or jurisdiction. 

If used, there must exist specialised legal and practical safeguards, which has yet to materialise in any members of the Council of Europe. Legal and practical safeguards refer to CSS being used fairly, a explicit public versus individual interest evaluation, but it cannot be the argument for why the Right to Remain Silent and Not Incriminate Oneself must be extinguished\footnote{54810/00, Jalloh v. Germany, § 97.} which has been firmly established by the ECtHR in various case law\footnote{34720/97, Case of Heaney and McGuiness v. Ireland, §§ 57 - 58.}. 

Essentially, public interest cannot extinguish the very essence of the right. CSS systems will likely require costly national implementation to not violate The Right to Remain Silent and Not Incriminate Oneself. CSAMD and similar systems have, as of the time of writing, not done so, and \hyperref[Section 5]{the proposed Regulation we briefly look at later} does not change this. Following this, we include further details within Article 6.

\subsubsection{Admissibility} The evidence must be admissible, which refers to whether evidence can be used at a trial or not. Some states have rules against evidence obtained wrongfully or by other means of evaluation, which can lead to situations where the files or just the detection of them cannot be used in proceedings \cite{Voeten2008}. The Convention does not consider these aspects, as these must be done in national law and by the relevant court\footnote{28490/95, Hulki Günes v. Turkey.}, but it evaluates whether the case was ``fair'' or not. The fairness test is individual and must be done entirely on a trial to trial basis. 

If the case is not deemed fair because of non-admissibility, some jurisdictions will close the current case, and others will continue but very likely lead to the accused being found not guilty. 

\subsubsection{Planted Evidence and Entrapment}\label{4.1.3} Relating explicitly to the Right to a Fair Trial, a trial can never be fair if the accused cannot question the evidence, which will be very difficult if relying on CSS or black box systems. This is due to the nature of the scanning constantly on the device of the accused, and because of the hidden or deliberately complicated or obfuscated approach to the coding. 

A trial cannot be fair either, if the evidence used was planted\footnote{22062/07, Layijov v. Azerbaijan, § 64.}. This can be done by an adversary, who sends the accused a file containing something that triggers the scanning. Regarding CSAMD, steganography \cite{Subramanian2021} is not likely to be able to pass NeuralHash, but for future CSS systems this may not case. Normally planted evidence refers to it being planted by the state, but this definition is expanded because of the increase possibilities of interference by third parties that have no physical access to the accused. Lastly, a trial cannot be fair if the evidence is obtained through unlawful secret surveillance\footnote{35394/97, Khan v. the United Kingdom, § 34.}, which as a general rule will make such evidence inadmissible.

Entrapment is worth mentioning in the same vein. This refers to various means to either frame or otherwise through ``traps'' to finds reasons to investigated individuals\footnote{59696/00, Khudobin v. Russia, § 128.}. This could very well be done with CSAMD or similar systems, with the exact same techniques as an adversarial attacker. Entrapment has happened in many states of the Council of Europe, including Croatia\footnote{47074/12, Grba v. Croatia.}, Germany\footnote{40495/15, Akbay and Others v. Germany.} Lithuania\footnote{74420/01, Ramanauskas v. Lithuania.}, United Kingdom\footnote{67537/01, Shannon v. the
United Kingdom} and more.

%Article 6 interacts with the following articles, so we will return to how it is relevant otherwise to CSAMD and similar systems.

\subsection{Right to Respect for Private and Family Life, Home and Correspondence}
\label{sec:privateliable}
Article 8 has to do with the integrity and status of one's private pictures. It includes the protection of privacy\footnote{40660/08 and 60641/08, Von Hannover v. Germany (no. 2), § 95.}. In the case law of the ECtHR, privacy is either protected positively or made to protect against the state negatively, but there exists a right for the individual to be protected against other private parties\footnote{61496/08, Bărbulescu v. Romania [GC], §§ 112 - 112, to be understood in an expanded manner.}. 

This right for protection against private parties is a recognition of their new role in infringing privacy systematically. This allows Article 8 to be applied horizontally under certain circumstances, which enables individuals to sue to parties which are not just the state, or at the very minimum sue the state over its lack of action against the private parties. 

We see that the Right to Privacy could potentially render the use of CSS illegal within national states. This comes down to the wording of Article 8, which in part 1, sets out the right and is qualified in part 2. The Right to Privacy is therefore not an absolute right, but instead relies on the entirety of Article 8. Interfering in privacy requires a legitimate aim, and ``fighting'' CSAM would fit under ``prevention of order or crime'', ``for the protection of health and morals'' and ``for the protection of the rights and freedoms of others'', usually not an issue to prove in court\footnote{43835/11, S.A.S. v. France [GC], § 114}, but other CSS systems might not pass. The latter could be because of their purpose not being related to criminal investigations. 

The other test from part 2 of Article 8 however, is whether violating the right is ``necessary in a democratic society''. This is by the court interpreted directly into ``pressing social need'', which the interference must justify precisely, and it must be proportionate to the problem which needs to be solved\footnote{20071/07, Piechowicz v. Poland, § 212}. Apple would then need to justify the surveillance through these rules, and for CSAMD, this has not been done adequately. There is no documentation that proves or makes it likely that there exists the huge amount of CSAM which would justify CSS on all devices, as the measure would only be proportionate if the pressing social need warranted the breach of privacy. There is a chance that such documentation can be brought to light in court proceedings. 

The same can be said about CSS for location data, e.g. contact tracing applications \cite{White2021}, which so far would not be able to justify its existence via part 2 either. Note that this can be done without invoking data protection legislation. This does not mean that the surveillance will be criminalized, but practice indicates that it could be possible\footnote{38435/13, B.V. and Others v. Croatia, § 151.}, even if civil litigation is more likely\footnote{25163/08, 2681/10 and 71872/13, Noveski v. the former Yugoslav Republic of Macedonia, § 61.}. 

Like many types of existing surveillance, the ECtHR is not willing to directly require the removal of the techniques, most clearly seen with the techniques of entrapment from \hyperref[4.1.3]{ Section 4.1.3} earlier, which are allowed narrowly\footnote{74420/01, Ramanauskas v. Lithuania [GC], § 51}.

%How come? How does mass surveillance software which makes use of precarious encryption techniques and which risks the privacy of the individual by its mere existence or lack of security violate Article 8?

\subsubsection{Violation by Existence} Questions could then arise as to how such surveillance systems could violate privacy of individuals by merely existing, such as through the medium. Depictions and pictures of individuals are inherently protected by Article 8\footnote{1874/13 and 8567/13, López Ribalda and Others v. Spain [GC], §§ 87-91.}. 

This means that the individual must be protected against state and other private actors intruding on this right\footnote{40660/08 and 60641/08, Von Hannover v. Germany, §§ 50-53.}. 

This is not absolute, and no case law that relates directly to the protection outright without any criterion\footnote{18068/11, Dupate v. Latvia, §§ 49-76.} exists. There is no case law which guarantees that any surveillance could be justified if sufficiently ``privacy enhanced''\footnote{30562/04 and 30566/04, S. and Marper v. U.K, § 125.}, as the court leaves no room for exceptions if the surveillance is disproportionate and violates any degree of appreciation the state had or fails the tests above. We narrowly interpret this as Article 8 never allowing privacy enhancing technologies to justify the measures. CSS systems like the CSAMD must therefore be analysed on the basis of protecting the depiction of the individual, since this is the subject of the surveillance. 

The question then becomes how far it reaches. Generally, the state must positively protect this right through criminal or civil law provisions\footnote{5786/08, Söderman v. Sweden.}. It could mean that to use CSS, there should be legal safeguards that force Apple or other companies to use state of the art encryption and demand local representatives that act as humans in the loop in their system. Other authors have suggested similar policy proposals \cite{Abelson2021}.

\subsubsection{Defamation} The protection of individual reputation is also included in Article 8. An attack of a certain level of seriousness must have occurred on the individual and it must have harmed the personal enjoyment for the right to respect private life\footnote{76639/11, Denisov v. Ukraine, § 112.}. This could be the planting of CSAM or false positives. 

If CSS systems ends up causing loss of reputation through the media or through companies/authorities themselves accusing the user, Article 8 can apply. The most important concept in this aspect is whether the loss of reputation was caused by user's foreseeable actions\footnote{25527/13, Vicent Del Campo v. Spain}, or if the loss of reputation was caused by a criminal conviction, which the court does not accept\footnote{76639/11, Denisov v. Ukraine [GC], § 98.}. 

From case law, this includes any part of the chain, and countries like the UK will allow this to be the basis for tort law or reimbursement law, with the company or state responsible as the potentially liable party \cite{Hughes2016}. %\cite{Hughes2016}. %This could be through false negatives in the CSAMD.

\subsubsection{State Surveillance} Because CSS systems may be part of secret state surveillance systems, similar safeguards as normal surveillance must be given\footnote{5029/71, Klass and Others v. Germany, § 36.}. These should include appeal processes and right to access. This is outside of what would be given through data protection rules, as access to it is required to receive the process of a fair trial. In an EU context, such data would not be covered by the GDPR, but instead the Law Enforcement Directive, which limits the potential information which the individual would be able to obtain to a limited amount, and which requires the surveillance to fulfill a range of criteria, including authorisation \cite{Rezende2020}. 

ECtHR practice shows that state surveillance via CSS without safeguards would be in violation of Article 8\footnote{4647/98, Peck v. the United Kingdom, § 59.}, but there is one issue with this approach. Getting a case before a national court or the ECtHR requires that the state surveillance is perceived or discovered. 

If the information provided by the client side scanning systems is the source of the secret surveillance, discovering it will be much harder because of how much further up it is up the surveillance chain. This is difficult, as it involves discovering and understanding the CSS which is embedded in systems used on a daily basis by those under surveillance.

\subsection{Freedom of Expression}

Article 10 applies to any medium\footnote{10572/83, Markt intern Verlag GmbH and Klaus Beermann v. Germany, § 26.}. CSS systems like the CSAMD will not immediately impact the Freedom of Expression of any individual. Derived (chilling) effects seen in other types of surveillance systems will regardless of the intention lead to diminished rights, especially freedom of speech \cite{Sen2002, Solove2011} or theoretically any type of freedom \cite{Schwartz1999}. 

\subsubsection{The three ``tests''} Like the other ECHR rights discussed, it is made in a positive manner. As the exception, there exists a test which the courts have to take regarding situations where Freedom of Speech can be suppressed \cite{Gunatilleke2021}. This consists of three parts, which must be cumulatively fulfilled for the state to be able to justify violating Freedom of Speech: \emph{Lawfulness of the interference (1), legitimacy of the aim pursued by the interference (2), necessity of the interference in a democratic society (3).} Failure at any stage will make the surveillance be in violation with Article 10. %\footnote{All three are expressed in art. 10, as; "prescribed by law", "pursued one of the legitimate aims" and "necessary in a democratic society".}

\emph{1.} Lawfulness of the surveillance system in interfering Freedom of Expression is attained with positively describing and implementing it in law, with the latter requiring precise wording and literal usage and meaning of the text\footnote{24973/15, Cangi v. Turkey, §§ 39 and 42.}. The Court recognises that technology changes, and that wording can be made to be vague, but usage matters, and in this sense ``what'' the surveillance actually does\footnote{21279/02 and 36448/02, Lindon, Otchakovsky-Laurens and July v. France, § 41.}. This catches the issue of lawfulness, that it must be foreseeable for the user or individual whose Freedom of Expression is infringed. 

Future systems require proper implementation into law to not threaten Freedom of Expression through picture recognition of people and so on. This rules out any kind of private surveillance, which states would then have to protect its citizens against, making private versions of such systems illegal. It is not foreseeable for the individual that all their storage or other digital devices suddenly become live CSS systems, at least not within current case law of the ECtHR.

\emph{2.} The surveillance must pursue a legitimate aim. This can be passed depending on the lack of constitutional protection of the system, or through popular movements, but they must be clear\footnote{67667/09 and others, Bayev and Others v. Russia, §§ 64 and 83.}.

\emph{3.} The third test is the most divisive, and the case law for this decides what should be done, not the wording as such. There must be a ``pressing social need''. Pressing here refers to an actual need, and may not be twisted in practice or other through legislation\footnote{6538/74, The Sunday Times v. the United Kingdom, § 59.}, though this does not make it indispensable. CSS could therefore be relevant during an insurrection or other emergencies or due to other extreme circumstances. Secondly, the assessment of the severity of the system must not lead to the assumption that it causes censoring\footnote{56925/08, Bédat v. Switzerland, § 79.}. Any kind of sanctions associated with the system must be proportional\footnote{13444/04, Glor v. Switzerland, § 94.} and properly legislated and justified\footnote{Bayev and Others v. Russia, supra, § 83.}. Thirdly, national courts can be part of the problem if they are not able to sufficiently and properly assess Article 10 in regards to this issue\footnote{23954/10, Uj v. Hungary, §§ 25-26.}. 

%This is rather unique, but makes sense, since this last bastion of justice can be perverted under certain circumstances. 

%\subsubsection{Hosting Data}
%Companies like Apple, Google and Microsoft currently host billions of pictures of individuals. We can argue that these are ``given'' in confidence to them, which means that further case law and considerations will apply. Considerations of whether the contract one signs when using these services are established and has legal effect, is for another paper, but they are unlikely to have the intended effect in most jurisdictions due to consumer and contract law. The question becomes when and under which circumstances can they relay the information they have gained from automatically analyzing personal pictures to authorities or other private parties. 

%intentional question, should be changed.

\subsection{Freedom of Assembly and Association}

%This kind of surveillance moves the purpose of assembly and association if it disables 

CSS can be used long term to control or infringe given Article 11 rights. This can be done through recognition of pictures, locations, people or specific subjects. The aims for this kind of surveillance could be suppression of political or non-political parties, specific public figures or unions \cite{Aston2017, Siatitsa2020}. 

Because of the nature of the surveillance, we focus on Freedom of Association. Article 11 and Article 10 are not in competition, rather the opposite, and Freedom of Assembly is clearly as important as Freedom of Expression\footnote{20652/92, Djavit An v. Turkey, 2003, § 56}. The freedom to associate with any political party is crucial\footnote{133/1996/752/951, United Communist Party of Turkey and Others. v. Turkey, 1998, § 25.} as is any other form of group\footnote{48848/07, Association Rhino and Others v. Switzerland, 2011 § 61.}. To be included in the protection, the 'association' must have a private character\footnote{7601/76 and 7806/77, Young, James and Webster v. the United Kingdom, 1979, Commission’s report, § 167.}, but the state cannot speculate in nationalising it on purpose to remove the Article 11 protection\footnote{42117/98,  Bollan v. the United Kingdom, 2000.}, or in reality prevent any ineffective exercise of the right\footnote{70945/11 et al., Magyar Keresztény Mennonita Egyház and Others v. Hungary, 2014, § 78.}. Article 11 states that any intervention with the right must ``prescribed by law'', pursue legitimate aims and be “necessary in a democratic society''. These are not defined in the same manner as Article 10. To be prescribed by law, the intervention of the Right of Assembly must be positively described in legislation, must be available for those affected and must be foreseeable\footnote{39748/98, Maestri v. Italy [GC], 2004, §25 - 42.}. 

Any CSS systems used for these purposes must be included in national legislation and the public must be clearly warned that they are able to use it to essentially prevent assembly. Preventing assembly could be in the form of dissolving organisations (political, labour or otherwise) through identification and arrests or milder economic sanctions. 

The core point is that the CSS would be used for identification, as pictures contain metadata be themselves or through the system, or clear location or other data, and that they can contain elements that could link the user to the organisation that the state wants to quash. CSAMD is likely capable of being re-purposing the system to look for these factors and through the same methods, alert a company (or the state using the system) which can then lead to further investigation and potentially litigate. Future systems are likely to be able to do this much easier and be designed for it.

\section{The Child Sexual Abuse Regulation} \label{Section 5}

The proposed EU Regulation laying down rules to prevent and combat child sexual abuse \cite{CSAregulationproposal} (CSA) has issues which we must comment on. In its impact assessment of the proposed legislation, under its analysis of loss of fundamental rights, the European Commission claims on page 14 claims that CSS is ``often the only possible way to detect it'', foregoing all other types of preventive and criminal measures Member States can take against CSA. 

There is no argumentation via logic or backed up by literature. This makes the statement and perhaps the whole Regulation techno-solutionist, a term which much good research \cite{Gardner2019} illustrates, which implies the argument merely want to only solve a problem with a specific technology without regarding other factors. The European Commission furthermore disregards and does not analyse the potential consequences either CSS or server-side scanning would have on cybersecurity and privacy, while they justify the victim's potential positive outcomes outweighing the negative of everyone else. The main tools of the Regulation are:

Providers must conduct risk assessments (Article 3) and providers must mitigate risks (Article 4). 

Force app stores to prevent children from using inappropriate (not well defined) apps (Article 6), and force CSS or server-side scanning for all providers of communication services (this includes Signal) (Article 10) and backdoors, combined with potentially creating unlimited preservation of data if requested (Article 22).

Enforcement powers, which includes 6 percent of annual turnover or global income based fine (Article 35)  and forcibly physically shutting servers down (Article 28 and 29).

The creation of a EU Centre to facilitate technology and support providers and Member States, and have databases that are not well specified, without any requirements for its own staff and no assurance that EUROPOL will not \emph{de facto} control or otherwise influence it.

We know very well that self-regulation requires teeth if not complied with, which the Regulation does have, but its requirements end up putting the entire sector at risk from how zealous it looks or through its attempts to solve its goals in Article 1 (impossible only with these tools, see \hyperref[CSS impossible]{Section 2}). 

Depending on the interpretation, Article 10 may compromise all communication platforms which are used by EU citizens, and the European Commission has not learned from the mistakes of the past regarding key escrow, nor have they taken heed from existing literature such as \cite{Abelson2021}.

\section{Future Considerations}

%Distinguish between adversarial failure and non-adversarial failures. 

If these systems get implemented into our lives through all digital infrastructures, keeping check on their influence and consequences and what can be done to use them is of utmost importance, which means increased amounts of research into their actions upon the lives of those afflicted by them, whether natural or legal persons. This research includes judicial, cybersecurity and structural reviews, as well as more empirically based research through interviews of those that develop or are affected by them. Plenty of existing research into other systems is usable \cite{Stoycheff2020, Mann2020}, but because of the character of these new surveillance structures, the reverberations of the actions of the system will be that much greater. 

We hope that there will be an increasing focus in the research on the downsides and or the costs of CSS, in essence the proportionality of the exercise. This question could be answered with an analysis focused on proportionality as a principle within law or philosophy, and would fit as an extension of existing research \cite{Cobbe2021, Rosenzweig2020}, and the development of the EU CSA Regulation. A continuation of showing the impact on human rights of every kind by CSS is needed as well.

\section{Conclusion}

%Sentence for later into the conclusion or in the start

%change conclusion in general, this is a temp thing

If you want to dig for gold, you predict accurately where it is. What you usually do not do, is to dig up the entire crust of the surface of the earth. CSS systems and mass surveillance represent the latter. 

In this paper, we argue that CSS cannot be justified to be used in the way systems like CSAMD is presented. We do so on the basis of two different kinds of proofs, but with the same outcome: Because CSS cannot reach its own purpose, either due to assumptions on universal circumventability or purely because it cannot technically or practically reach the goals, it will not be fit for fighting CSAM by itself. 

We are aware of ongoing developments such as Encrochat , but that specific case represents a situation where the entire infrastructure was compromised, and was more of a trap than a CSS. There are far more elaborate and safer alternatives than the PSI and surrounding system which Apple chose for CSAMD, and this could in retrospect be considered the first major ``failure'' of the system \cite{Abelson2021}. Instead of relying on these safer alternatives, Apple decided to create an infrastructure which beckons to suffer adversarial failures from the start \cite{Struppek2021}. 

While this may change, this does not prevent the second ``failure'' of the system. We show in this paper that CSS like CSAMD will have the potential to not just violate one human right, but most likely $ \leqq 2$ rights at the same time. CSS such as CSAMD is very likely to violate the EHRC Article 6 derived Right to Remain Silent and Not Incriminate Oneself, which could cause serious admissibility issues in courts, potentially preventing trials against criminals or otherwise disrupt legal systems. CSS opens endless possibilities for states to use entrapment, potentially causing the prosecution of innocent individuals. 

Another violated right is Article 8, which includes a right and protection of privacy. CSS will not pass the test of "necessary in a democratic society", exactly because these systems rarely will be proportionate in their infringements to their goals. We discussed two more potential violations, which CSS may cause, but these were both hypothetical. This was Article 10, Freedom of Expression, and Article 11, Freedom of Assembly and Association. We note that CSS systems would rarely meet the thresholds to justify violating the rights, but what is important to acknowledge is the very powerful ways which they could already do so. Surveillance systems seen in China are likely already capable of this \cite{CorinneReichert2019, Laskai2021}. 

We comment on the proposal for a Regulation by the European Union to fight Child Sexual Abuse, which worryingly it does through mandating CSS and breaking the encryption of all messaging services which can be identified as such. We show in our proofs that CSS is by no means the right or adequate solution, and the proposal does not justify how this could be achieved without fundamentally breaking everyone's right to privacy and fair trial.

%Finally, we make some overarching assertions regarding the role which CSS may have in criminal large in general, which, combined with the problems which the manufacturers of these systems may face themselves. In that sense, CSS is turning into another surveillance system on top of everything else.

%Lawful authorisation, unregulated or nontransparent cooperation with authorities and the change of parameters to suit different needs rather than fight CSAM, allow such systems to violate both e.g., Right to Fair Trial and Freedom of Expression. Our list and analysis does not cover all rights which may be violated by similar systems in the future, but the conclusions shown should be a cause of concern, as it may have a chilling effect on the lives of those which are under surveillance, and this is just within the context of Europe and those that obey the ECHR.

%end sentence, not findings here, add above

\printbibliography

\end{document}